# Association between Air Pollutants and Hospitalizations for Cardiovascular Diseases: Time-Series Analysis in São Paulo, 2010–2019

**Carlos Souto dos Santos Filho***

19060-900, Presidente Prudente - SP, Brazil
carlos.souto@unesp.br

**Ana Júlia Alves Câmara**

29.075-910, Vitória- ES, Brazil
ana.j.camara@ufes.br

**Guilherme Aparecido Santos Aguilar**

19060-900, Presidente Prudente - SP, Brazil
guilherme.aguilar@unesp.br

September 15, 2025

## Abstract

Cardiovascular diseases (CVD) remain one of the leading causes of hospitalization in Brazil. Exposure to air pollutants such as PM10µm, $NO_2$, and $SO_2$ has been associated with the worsening of these diseases, especially in urban areas. This study evaluated the association between the daily concentration of these pollutants and daily hospitalizations for acute myocardial infarction and cerebrovascular diseases in São Paulo (2010–2019), using generalized additive models with a lag of 0 to 4 days. Two approaches for choosing the degrees of freedom in temporal smoothing were compared: based on pollutant prediction and based on outcome prediction (hospitalizations). Data were obtained from official government databases. The modeling used the quasi-Poisson family in R software (v. 4.4.0). Models with exposure-based smoothing generated more consistent estimates. For PM10µm, the cumulative risk estimate for exposure was 1.08%, while for hospitalization, it was 1.20%. For $NO_2$, the estimated risk was 1.47% (exposure) versus 1.33% (hospitalization). For $SO_2$, a striking difference was observed: 7.66% (exposure) versus 14.31% (hospitalization). The significant lags were on days 0, 1, and 2. The results show that smoothing based on outcome prediction can generate bias, masking the true effect of pollutants. The appropriate choice of df in the smoothing function is crucial. Smoothing by the pollutant series was more robust and accurate, contributing to methodological improvements in time-series studies and reinforcing the importance of public policies for pollution control.

Introduction

The effects of exposure to environmental factors on chronic degenerative diseases are the most widely investigated in environmental toxicology. Several epidemiological studies have reported that a combination of environmental risk factors increases the likelihood of CVD events and deaths[1,2,3].

Cumulative evidence on air pollution published by the American Heart Association supports the causal relationship between exposure to particulate matter smaller than 2.5 µm (PM2.5µm) and cardiovascular morbidity and mortality[4]. Other previous studies have reported that exposure to PM2.5µm causes approximately 70% to 80% of premature cardiovascular deaths[5]. Environmental pollutants such as volatile organic chemicals, Carbon Monoxide (CO), Nitrogen Dioxide ($NO_2$), Sulfur Dioxide ($SO_2$), Ozone ($O_3$), and sulfates have been reported by experts from the Council on Epidemiology and Prevention of the American Heart Association as contributing factors to the risk burden of CVD, as these pollutants constitute about 98% of the gas mixture of urban centers and can optimize the harmful effects of PM[6].

Nationally, researchers investigated the relationship between daily mortality among elderly individuals over 65 years of age and environmental pollutants in the metropolitan region of São Paulo. Using time-series models, a strong association between mortality and particulate matter <10µm (PM10µm), nitrogen oxides ($NO_x$), sulfur dioxide ($SO_2$), and CO was found. The association with PM10µm was the most significant, resulting in a 13% increase in overall mortality[7].

Starting in the 2000s, research on environmental exposure to air pollutants and hospitalizations for CVD among elderly individuals in the city of São Paulo was published, with an emphasis on generalized additive Poisson regression models adjusted for lagged effects up to 20 days after exposure. The pollutants PM10µm and $SO_2$ were associated with a 3.17% increased relative risk (RR) for congestive heart failure and 0.89% for total cardiovascular disease. Authors have commented on the strong association found between exposure to air pollutants and cardiovascular hospitalizations[8].

Based on the studies described above, empirical evidence of the impact of environmental exposures to atmospheric toxic agents and meteorological conditions on human health is a subject of global relevance[9]. These studies serve as a reminder of the importance of controlling and reducing atmospheric pollutant emissions and supporting health managers in resource allocation.

Interest in this topic has grown exponentially, with publications also focusing on quantifying and monetizing the effects of covariates (air pollutants and climate) on various economic sectors, environmental resources, energy consumption, air quality, and human health. Meta-analytic research on the economic impact of these covariates on health has linked average temperatures to mortality rates from respiratory and cardiovascular diseases[10]. Patz et al.[11] conducted predictive studies of increased health risks under future climate change projections. With the results obtained, the authors also pointed to increases in morbidity and mortality on a global scale.

This research is aligned with the UN 2030 Agenda, especially SDGs 3 ("Ensure healthy lives...") and 13 ("Combat climate change...")[12], and reaffirms the institution's commitment to sustainable development. In this context, it is essential not only to apply but also to improve methodologies capable of accurately capturing the effects of environmental exposures on health.

A niche of methodological issues related to the analysis of the relationship between air pollution and health has attracted the attention of the scientific community, namely time-series and semiparametric models, which are the standard approaches for studying short-term associations between daily exposures (e.g., air pollution, temperature) and health outcomes (e.g., mortality or cardiovascular hospitalizations). Data are usually collected at regular intervals (e.g., daily) and models aim to estimate exposure-response relationships, adjusting for temporal trends and confounding factors[13,14].

Most previous studies in Environmental Epidemiology and Time Series are based on generalized linear models with a logarithmic link and Poisson error, regressing daily counts on air pollutant levels and covariates. Day-of-the-week and meteorological terms (e.g., temperature and humidity) are included as covariates, and smooth functions of calendar time are used to control for long-term trends and seasonality. This semiparametric Poisson regression framework produces estimates of the relative logarithmic rate (risk) per unit of exposure change[14].

According to Tomov et al.[15], unlike forecasting models, which aim to project future outcomes, time series models applied to the environmental context focus on estimating the effects of exposure, seeking to explain observed data, rather than predict future events.

To control seasonal and long-term trends in time series, a widely adopted strategy is the use of semiparametric models that incorporate smooth functions of time. The literature highlights the use of GAMs with Poisson distribution and nonparametric smoothing—such as LOESS—to adjust variables such as time, temperature, humidity, and pollutant concentrations, especially PM10µm[16]. In this approach, smooth functions act as filters that remove seasonal patterns and persistent trends, allowing for a more accurate estimate of the associations of interest. Several studies support the use of flexible representations of these functions, using techniques such as splines (smoothing, penalized, and natural)[17,18]. It is important to emphasize that, although such adjustments consider long-term temporal variations, factors with faster fluctuations can also act as potential confounders and require appropriate control in the model.

The use of smooth functions of time in semiparametric models raises important statistical questions, especially due to the uncertainty regarding the shape and complexity of seasonal and long-term trends in mortality and pollution time series. One of the main methodological difficulties is defining the appropriate degree of smoothing to apply. This choice is crucial, as it directly impacts the amount of residual temporal variation that can be attributed to the effect of air pollution exposure. An overly smooth adjustment can eliminate relevant variations and underestimate the true effect of the pollutant, while insufficient smoothing can leave residual temporal patterns that introduce bias due to confounding. Defining the level of smoothness has been addressed through automatic methods guided by adjustment criteria (such as the information criterion) or by incorporating prior knowledge about the temporal scale of potential confounders. However, these methodological decisions still lack clear guidelines and represent a significant challenge in statistical modeling applied to environmental health[19].

This study aims to investigate the association between daily exposure to PM10µm, $SO_2$, and $NO_2$ and the relative risk (RR) of CVD hospitalizations in the city of São Paulo, Brazil. In addition to filling this gap, the research seeks to improve the methodological approach through semiparametric time series models that incorporate smooth functions of time. This study raises the question of whether the appropriate definition of the degree of smoothing directly impacts the reliability of the results.

In this study, we propose a comprehensive evaluation of model specification strategies in environmental time series analyses, focusing on the choice and impact of the df of the temporal smoothing function in estimating the effects of air pollutants on CVD hospitalizations. First, we compare analytical approaches commonly employed in environmental epidemiology to model long-term and seasonal trends, contrasting two distinct criteria for selecting df: (i) based on the prediction of the exposure variable (pollutant levels) and (ii) based on the prediction of the health outcome (hospitalizations for CVD).

We then applied these approaches to a database composed of daily air pollution data (PM10μm, $NO_2$, and $SO_2$), meteorological variables, and records of hospital admissions for ischemic heart disease and cerebrovascular disease in the city of São Paulo, from 2010 to 2019. The modeling was conducted using GAM with multiple lags (from 0 to 4 days), following the approach adopted in previous studies to capture delayed effects of exposure[20], and a quasi-Poisson distribution, aiming to estimate the short-term impacts of pollutant exposure.

Finally, we sought to quantify the magnitude of the bias introduced by the different temporal smoothing choices, assessing their implications for risk estimates and the associated uncertainty.

## Methods

### Study Area

The city of São Paulo belongs to a regional political and socioeconomic division of the state of São Paulo, in southeastern Brazil. The municipality has a land area of 1,521.202 km² and an urbanized area of 914.56 km² (as of 2019). It has a population density of 7,528.26 inhabitants per square kilometer and approximately 11,451,999 inhabitants (as of 2022). The municipality's industrial and economic sector is robust, with a per capita GDP of R$66,872.84 (as of 2021). In 2023, it ranked first in total revenue and total expenditures. Vehicle emissions are one of the predominant sources of pollution. In August 2024, the municipality had a total fleet of 9,721,123 vehicles, of which 6,353,930 were cars and 146,701 were trucks. Its boundaries are given by the Tropic of Capricorn and it is located between latitudes 23°20' and 24°00' S and longitudes 46°20' and 46°50' W, at 760 meters above sea level[21]. According to the Köppen-Geiger classification[22], São Paulo's climate is subtropical Cfa. Its territorial characteristics allow for a transitional climate between humid highland tropical climates and permanently humid subtropical climates, demarcating a hot and humid summer season and a cold and dry winter season.

### Data Extraction

The health information used in this study was obtained from digitized and anonymized data from the municipality of São Paulo, provided by the DATASUS. The dependent variable (DV) corresponds to the daily number of hospital admissions for ischemic heart disease and cerebrovascular disease, based on records from the Hospital Admissions System between 2010 and 2019. The conditions were classified according to ICD-10 codes: I21, I24, I50, I60 to I69. For analytical purposes, data were stratified by age group (<65 years, 65 to 75 years, and >75 years) and gender (male and female). These categories were incorporated as covariates in the statistical model to adjust estimates and assess potential effect modifications in the associations between pollutants and cardiovascular outcomes. Racial, educational, and occupational criteria were not studied.

Air pollutant data (PM10µm, NO2, SO2) were obtained from the São Paulo State Environmental Agency for the following monitoring stations: Santana, Parque Dom Pedro II, Congonhas, Ibirapuera, Mooca, Cerqueira Cesar, Cidade Universitária USP-Ipen, Nossa Senhora do Ó, Itaquera, Interlagos, Itaim Paulista, Marginal Tiete-Pte Remédios, and Perus.

The climatic (confounding) covariates were extracted from the meteorological database of the National Institute of Meteorology. The covariates (air pollutants and climate) were extracted from the same time period as the DV.

*Computational Statistical Analysis*

As a starting point, an exploratory data analysis was performed. This included searching for outliers and missing data, as well as possible data imputation. Next, the covariates were selected to compose the final analysis. Age stratification into <65 years, 65–75 years, and >75 years was adopted according to criteria established in another study (Peng; Dominici; Louis, 2006) and due to physiological changes related to aging. Furthermore, the time series was decomposed into long-term trends, seasonal trends, and higher-frequency short-term trends (classical time series decomposition) for DV and covariates. R statistical software (version 4.4.0) and the following packages (dyn, broom, stargazer, quantmod, mgcv, dlnm, stats, tsModel, dplyr, splines, gam, rempsyc, dyn, lattice, mda) were used. This study was reported in accordance with the STROBE guidelines for observational studies.

*Semiparametric Time Series Model*

In the scenario where $Y_t$, $x_t$ e $z_t$ (climate variables) and unmeasured potentials $s(t, \lambda)_t$, semiparametric models are an attractive methodological framework. These hybrid models combine the advantages of parametric and nonparametric models by allowing the inclusion of explicit parametric terms for specific predictors (e.g., PM) and smoothed nonparametric terms for other predictors.

Therefore, for this research, GAM was applied in combination with a nonlinear distributed lag model, a modeling framework that can simultaneously represent nonlinear exposure-response dependencies and delayed effects[23,24]. This adaptation of the GAM is due to the fact that the DV exhibits temporal correlation. The conceptual expression of the modified GAM is:

$$\log \mu_t = \alpha + \beta x_{t-\ell} + s(z_t, \lambda_1) + s(t, \lambda_2) + \varepsilon_t$$

where $Y_t$, is a time series of daily counts of hospitalizations for CVD ($t = 1, \dots n$), $\log \mu_t$ is a logarithmic link function, $\alpha$ is the intercept, $x_{t-\ell}$ a daily time series of PM10µm, $NO_2$ and $SO_2$, and $z_t$ a time series of daily average temperatures, $s(z_t, \lambda_1), s(t, \lambda_2)$ are used to indicate a smooth function of $z_t$ and time $t$ respectively (proxy variables). It is observed that $z_t$ (e.g., mean temperature) has a smooth but unspecified relationship with hospital admission for CVD. The smoothness is controlled by the $\lambda_1$ df.

A distributed lag model was used, where multiple air pollution lags are entered simultaneously. It has the following conceptual framework:

$$\log \mu_t = \alpha + \sum_{\ell=0}^{K} \beta x_{t-\ell} + other\ factors\ t + \varepsilon_t$$

where $K$ is the maximum lag. In the specialized literature, the values of $K$ varies from 2 to 40 days[25].

The RR with a 95% confidence interval for hospitalizations due to CVD, stratified by age group and gender, was calculated for significant variables (p-value < 0.05), taking into account the cumulative effect of exposure (lag). Finally, in the final step, the best model was selected and its performance verified.

*Statistical Procedures: Smoothness Selection and df*

Time trend modeling was conducted using smooth functions to adjust for potential seasonal and long-term confounding factors in the estimates of the relative logarithmic risk associated with exposure to pollutants (PM10µm, $NO_2$, $SO_2$). To this end, the influence of the df used in the smoothing function was explored, given that different amounts of smoothing can substantially alter the estimated results.

To determine the appropriate amount of smoothing, an objective criterion was considered to avoid overfitting or underfitting the time function. Two main methodological approaches were evaluated:

Methods Based on Outcome Prediction: this approach selects the optimal number of df based on the model's ability to predict the outcome variable (CV hospitalizations). These methods are based on minimizing the Akaike Information Criterion (AIC) and the Bayesian Information Criterion, or minimizing residual autocorrelation using the partial autocorrelation function and white noise tests.

Methods Based on Exposure Prediction: the second approach is based on the generalized cross-validation (GCV) score or AIC minimization applied to models whose structure aims to predict exposure variables (daily levels of PM10µm, $NO_2$, $SO_2$). This strategy has the advantage of producing unbiased or asymptotically unbiased estimates of the association between exposure and outcome, which is particularly relevant in environmental studies[26].

Although the methods in the first category are widely used, their application in air pollution studies can result in biased estimates, as they optimize an adjustment criterion focused on the outcome rather than the exposure. Given the central objective of accurately estimating the risk associated with environmental exposure, we chose to compare both methods to verify the df selection strategy in temporal smoothing.

The statistical model adjusted using the gam() function of the mgcv package in R, considering a quasi-Poisson distribution to model the daily number of hospitalizations for CVD and the pollutant PM10µm, was as follows:

$$\log(\mathbb{E}[Y_t]) = \sum_{l=0}^{4} \beta_l \cdot \text{MP10}_{t-l} + f_1(days, \text{MP10}_t) + f_2(\text{average temperature }_t) + \varepsilon_t$$

where $Y_t$ the number of CV hospitalizations on the day t, $\text{PM10}_{t-l}$ is the concentration of PM10µm on the day $t-l$, for lags $l = 0,1,2,3,4$, $f_1(\cdot)$ is the smooth function for the time trend (based on the variable MP10µm), $f_2(\cdot)$ is the smooth function for compensated mean temperature and $\varepsilon_t$ are the residuals.

The modeling considers the lagged effects of exposure to air pollutants (PM10µm, $NO_2$, and $SO_2$), controlling for time trends and temperature. Smooth functions were specified with previously defined df, according to the methodological criteria described above. The RR was calculated using the equation:

$$RR = exp^{(\beta_1 * \Delta X)}$$

where, $\Delta X$ is the increment in exposure (e.g., 10 µg/m³ for each pollutant). The percentage variation in RR was given by the equation:

$$\% \ variation \ RR = \left(exp^{(\beta_1 * \Delta X)} - 1\right) * 100$$

This is a transformation commonly used in epidemiology to interpret effect sizes from log-linear models.

## Results and Discussion

From 2010 to 2019, 362,474 hospitalizations for ischemic heart and cerebrovascular diseases were recorded in São Paulo, representing a daily average of 99.25 hospital admissions. The gender distribution showed a male predominance, with approximately 55.83% of admissions (n = 202,377). Analysis of the descriptive statistics of hospitalizations revealed considerable variability in the data, with values ranging from 1 to 457 daily admissions. The first quartile (Q1) had 80 daily admissions, while the third quartile (Q3) reached 116 admissions, indicating a concentration of data close to the mean. The median of 100 daily admissions, close to the mean of 99.25, suggests a relatively symmetrical distribution of the central data.

Stratification by gender and age revealed distinct patterns of hospitalization. Among men, the age group with the fewest hospitalizations was those over 75 years old (average of 9.59 daily hospitalizations), in contrast to the male group under 65 years old, which had the highest average (31.72 daily hospitalizations). For women, a more balanced distribution was observed across age groups, with the over 75 group (12.66 daily hospitalizations) standing out, numerically outnumbering men in the same age group.

Monitoring of air pollutants revealed average concentrations of PM10µm of 26.02 µg/m³, $NO_2$ of 38.39 µg/m³, and $SO_2$ of 2.62 µg/m³. PM10µm variability was substantial, with maximum values reaching 110.12 µg/m³, much higher than the average. $NO_2$ showed a wider distribution, with maximum concentrations of 126.92 µg/m³, while $SO_2$ remained at relatively low levels, with a maximum of 18.79 µg/m³.

Meteorological conditions were characterized by an average temperature of 20.72°C, with a considerable temperature range between minimum (average of 16.71°C) and maximum (average of 26.41°C). Relative humidity averaged 73.10%, indicating moderate to high humidity conditions. Average total precipitation was 4.45 mm, with high variability represented by the high standard deviation.

Detailed analysis of the time series (Table 1) revealed interesting data characteristics. For hospitalizations due to CVD, the coefficient of variation of 0.26 indicated moderate variability. Positive skewness (1.63) and high kurtosis (21.49) suggested a distribution with a heavier right tail than normal and a greater concentration of extreme values. Air pollutants showed distinct variability, with $SO_2$ exhibiting the highest coefficient of variation (0.82), followed by PM10µm (0.59) and $NO_2$ (0.51). All meteorological variables demonstrated non-normal distributions, with precipitation exhibiting the highest variability (coefficient of variation of 2.58).

Table 1 – Time Series Statistics (Cardiovascular Disease Hospitalizations, Air Pollutants, Meteorological Data).

| Variable | SD | *Skewness* | *Kurtosis* | *Variance* | *Covariance* | *Coefficient of Variation* |
|---|---|---|---|---|---|---|
| **CVD hospitalizations** | | | | | | |
| Total CVD hospitalizations | 26.069 | 1.6290 | 21.490 | 679.608 | 679.608 | 0.262 |
| Male | 17.294 | 2.984 | 42.503 | 299.096 | 299.096 | 0.312 |
| Female | 11.549 | 1.059 | 14.641 | 133.387 | 133.387 | 0.263 |
| Male <65 years | 10.446 | 1.727 | 21.661 | 109.139 | 109.139 | 0.329 |
| Male 65–75 years | 6.722 | 4.337 | 44.856 | 45.189 | 45.189 | 0.476 |
| Male >75 years | 4.113 | 2.432 | 29.198 | 16.9171 | 16.917 | 0.429 |
| Female <65 years | 6.723 | 0.801 | 8.233 | 45.198 | 45.198 | 0.328 |
| Female 65–75 years | 4.044 | 0.785 | 8.004 | 16.360 | 16.360 | 0.378 |
| Female >75 years | 4.806 | 3.388 | 55.137 | 23.103 | 23.103 | 0.379 |
| **Air Pollutants** | | | | | | |
| MP10$\mu m$ | 15.320 | 0.795 | 4.397 | 234.703 | 234.703 | 0.588 |
| $NO_2$ | 19.748 | 0.379 | 3.527 | 389.997 | 389.997 | 0.514 |
| $SO_2$ | 2.152 | 1.455 | 6.643 | 4.634 | 4.634 | 0.820 |
| **Meteorological Variables** | | | | | | |
| Total Solar Radiation | 3.489 | -0.199 | 1.845 | 12.179 | 12.179 | 0.644 |
| Maximum Temperature | 4.385 | -0.322 | 2.657 | 19.236 | 19.236 | 0.166 |
| Mean Temperature | 3.432 | -0.279 | 2.760 | 11.779 | 11.779 | 0.165 |
| Minimum Temperature | 3.214 | -0.392 | 2.785 | 10.330 | 10.330 | 0.192 |
| Relative Humidity | 11.751 | -0.709 | 3.403 | 138.108 | 138.108 | 0.160 |

**Source:** Prepared by the author. **Note:** SD = Standard Deviation

Below is a graphical analysis of all the most relevant time series from this study (Figures 1 and 2). The analysis of the temporal variability of the pollutant PM10µm reveals a reduction in concentration variability over the 10-year period. After approximately 2015, there was a decrease in the frequency of extremely high PM10µm values, contrasting with the pattern observed in the previous period. This temporal shift suggests a possible change in environmental conditions or emission sources during the study period. It is worth noting that the graphical representation of the PM10µm data incorporates a temporal trend, allowing a clearer visualization of long-term patterns and the evolution of variability over the time series.

Figure 1 – Time Series of CVD Hospitalization, São Paulo, Brazil (2010-2019).

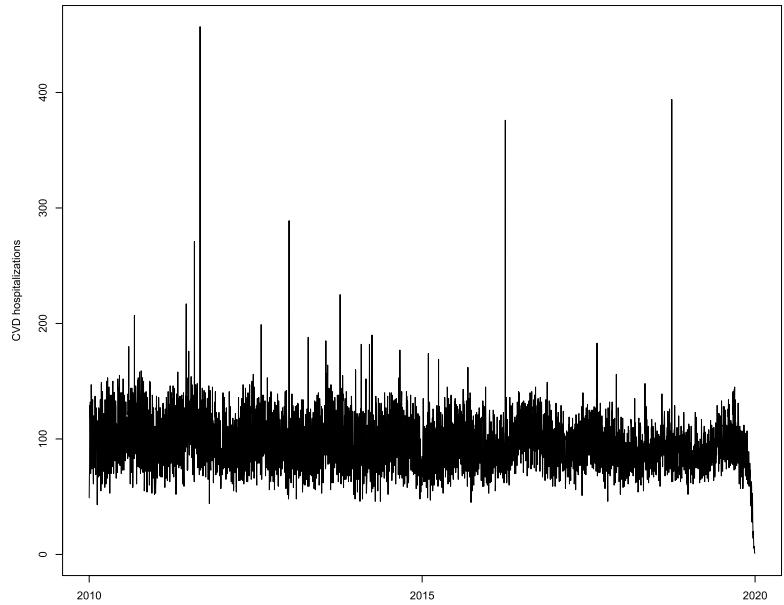

Source: Prepared by the author

Figure 2 – Time Series Graphs (Air Pollutants), São Paulo, Brazil (2010-2019).

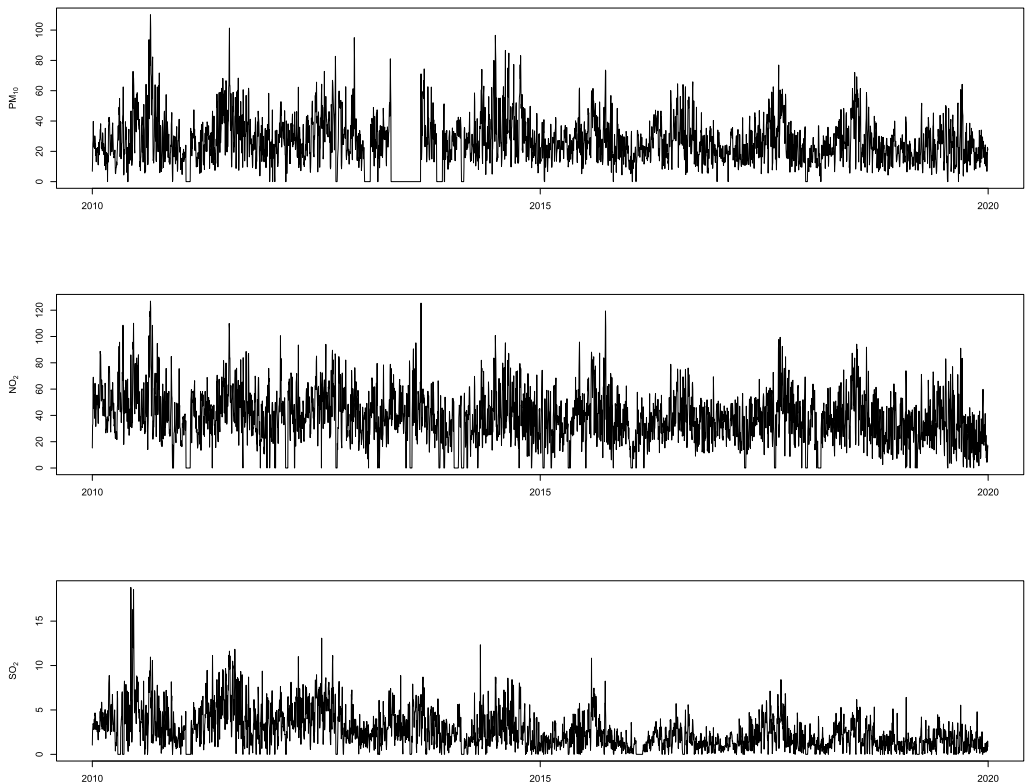

Source: Prepared by the author

After graphically analyzing the temporal trend of PM10 (Figure 2), a simple linear regression was applied. The model revealed a significant negative slope (Estimate = –0.0014, $p < 0.001$), indicating a consistent downward linear trend in PM10 concentrations over the study period. The intercept (Estimate = 48.50, $p < 0.001$) suggests that, at the beginning of the series, mean PM10 levels were approximately 48.5 µg/m³, reinforcing the evidence of a gradual reduction in concentrations across time. Applying the natural spline technique to the PM10µm data, using two degrees of freedom per year to capture seasonality, revealed a repetitive annual cycle characterized by peaks during the winter months and troughs during the summer months, influenced by meteorological factors and human emissions.

The CVD hospitalization data contain some days with extremely high counts. Therefore, these outliers were identified and removed for the next stages of the research.

Next, three time scales were considered for analyzing the above series: a single cycle (long-term trend), 2 to 14 cycles (seasonal trends), and 15 or more cycles (short-term and high-frequency trends), allowing for the identification of patterns across different time horizons (Figure 3).

Figure 3 – CVD Hospitalization Series on Three Time Scales, 2010-2019

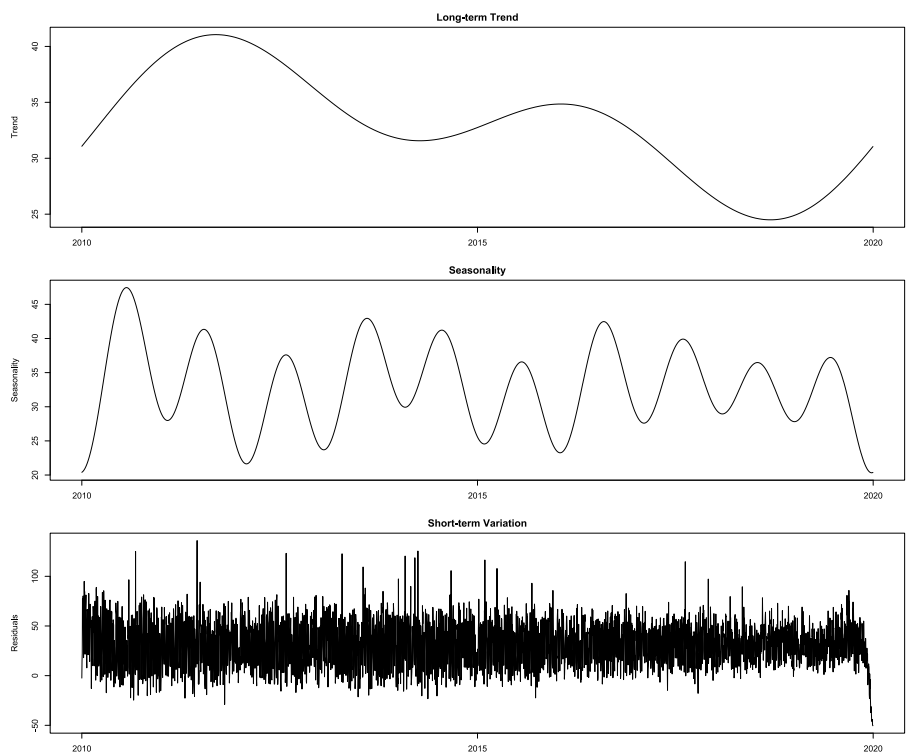

Source: Prepared by the author

The same procedure was applied to the predictor series (MP10µm) without outlier removal (Figure 4).

Figure 4 – Original MP10µm series at three time scales, 2010–2019

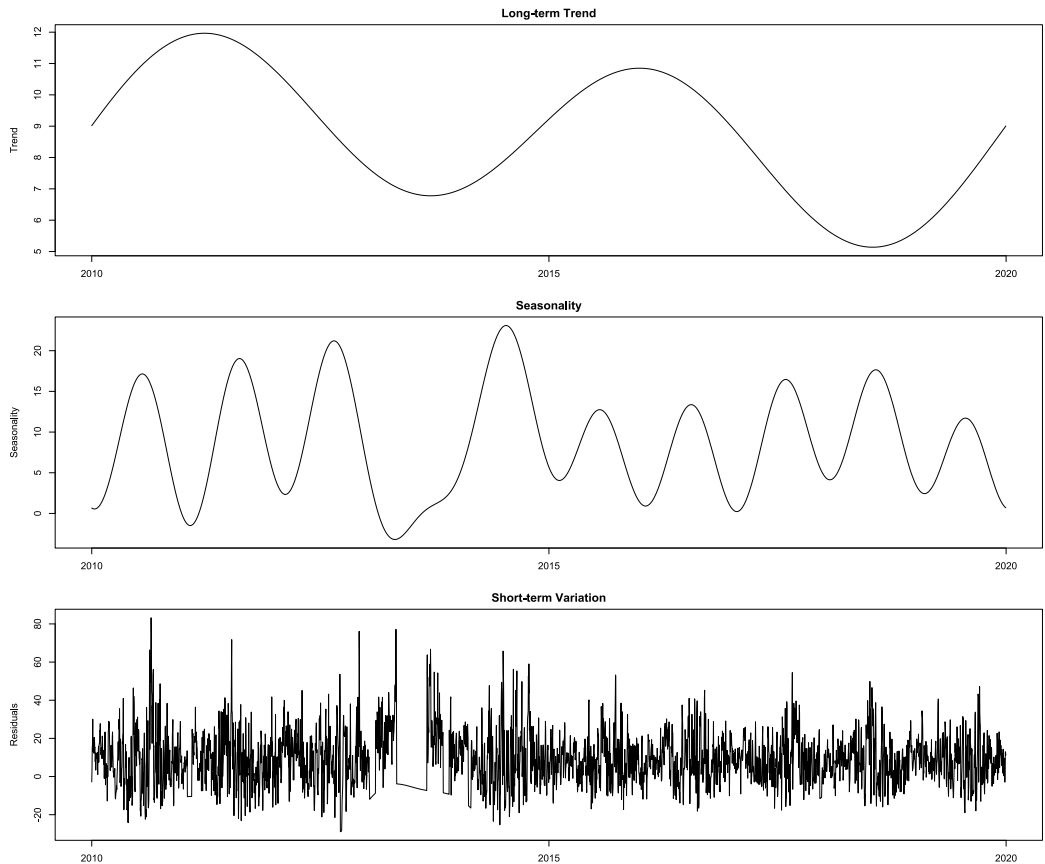

Source: Prepared by the author

Additional results are available as supplementary material accompanying the online article. Table 2 shows a comparison of RR values for the pollutants PM10µm, $NO_2$, and $SO_2$, which vary depending on the approach adopted to select the df of the smoothing functions. Models based on outcome prediction tend to underestimate the variability of the hospitalization series, resulting in greater dispersion and introducing bias into the estimate of the association between exposure and outcome. On the other hand, models based on exposure prediction presented more stable and, theoretically, less biased RRs due to their focus on minimizing the structural dependence of pollutants. These models, by prioritizing criteria such as GCV and AIC applied to exposure, produce more appropriate estimates, favoring robust interpretations of the relationship between air pollutants and hospitalizations for CVD. Table 3 shows a summary of the estimates of the coefficients of the pollutants PM10µm, NO$_2$ and SO$_2$ for the lags of 0 to 4 days, with the statistical significance represented by symbols according to the p-value ($\alpha = 0,05$).

Table 2 - Results of Estimates, Standard Error and DF by Year for Pollutants and Hospitalizations

| Pollutant | Model | Estimate | Standard Error | df / year |
|---|---|---|---|---|
| SO$_2$ | SO$_2$ Predictor | 7.66 | 0.0073 | 11.46 |
|  | Hospitalization Predictor | 14.31 | 0.0072 | 3.07 |
| NO$_2$ | NO$_2$ Predictor | 1.47 | 0.0007 | 13.58 |
|  | Hospitalization Predictor | 1.33 | 0.0007 | 3.07 |
| PM10µm | PM10µm Predictor | 1.08 | 0.0011 | 13.58 |
|  | Hospitalization Predictor | 1.2 | 0.0011 | 3.07 |

**Source: Prepared by the author**

Table 3. Lag coefficient estimates ($\hat{\beta}$) and significance levels

| MP10 | | | | |
|---|---|---|---|---|
| Lag | Estimate ($\hat{\beta}$) (Model 1) | p-value | Estimate ($\hat{\beta}$) (Model 2) | p-value |
| 0 | 0.00429 | < 0.001 (***) | 0.00445 | < 0.001(***) |
| 1 | -0.00155 | 0.0017 (**) | -0.00148 | 0.0032 (**) |
| 2 | -0.00111 | 0.0253 (*) | -0.00116 | 0.0210 (*) |
| 3 | -0.00004 | 0.9323 ( ) | -0.00002 | 0.9671 ( ) |
| 4 | -0.00050 | 0.1867 ( ) | -0.00059 | 0.1289 (.) |
| NO$_2$ | | | | |
| Lag | Estimate ($\hat{\beta}$) (Model 1) | p-value | Estimate ($\hat{\beta}$) (Model 2) | p-value |
| 0 | 0.00533 | < 0.001 (***) | 0.00533 | < 0.001 (***) |
| 1 | -0.00187 | < 0.001 (***) | -0.00186 | < 0.001 (***) |
| 2 | -0.00117 | 0.0004 (***) | -0.00122 | 0.0003 (***) |
| 3 | -0.00058 | 0.0777 (.) | -0.00059 | 0.0759 (.) |
| 4 | -0.00026 | 0.3241 ( ) | -0.00034 | 0.2013 ( ) |
| SO$_2$ | | | | |
| Lag | Estimate ($\hat{\beta}$) (Model 1) | p-value | Estimate ($\hat{\beta}$) (Model 2) | p-value |
| 0 | 0.02269 | < 0.001 (***) | 0.02453 | < 0.001 (***) |
| 1 | -0.00604 | 0.0611 (.) | -0.00447 | 0.1709 ( ) |
| 2 | -0.00819 | 0.0117 (*) | -0.00721 | 0.0282 (*) |
| 3 | 0.00136 | 0.6740 ( ) | 0.00188 | 0.5664 ( ) |
| 4 | -0.00244 | 0.3665 ( ) | -0.00135 | 0.6205 ( ) |

**Significance notes:** (***) p < 0.001, (**) p < 0.01, (*) p < 0.05, (.) p < 0.1, (vazio) p ≥ 0.1
**Source: Prepared by the author**

All quasi-Poisson MAG(s) models exhibited dispersion values well above 1, confirming the presence of overdispersion in the hospitalization count data and supporting the use of this modeling approach. Dispersion estimates ranged from approximately 5.15 for NO$_2$ under the exposure specification to 5.76 for SO$_2$ under the hospitalization specification. In general, $NO_2$ models showed the lowest dispersion values, while $SO_2$ presented the highest. Moreover, for each pollutant, the hospitalization specification yielded slightly higher dispersion estimates compared to the exposure specification, suggesting that temporal smoothing based on hospitalization degrees of freedom may have explained somewhat less variance than pollutant-specific smoothing splines (Table 3).

This study was designed to compare how different model specification strategies, particularly the choice of df in the temporal smoothing function, influence estimates of the effects of air pollutants on hospitalizations for CVD in analyses of ecological time series. The comparison between approaches based on exposure prediction versus those based on outcome prediction allowed a critical evaluation of the analytical practices commonly adopted in Environmental Epidemiology. The results reinforce the importance of aligning methodological choices with accurate estimation of the risk associated with exposure, highlighting potential implications of df selection on the validity of estimates.

Peng et al.[19] highlighted the limitations of prediction methods aimed at predicting outcomes, as they are developed to optimize an inadequate criterion. The authors argue that, in these studies, the primary focus is not on predicting health outcomes, but rather on precisely estimating the association between rising air pollutant levels and their health impacts. Therefore, methods that seek the best predictive model for the outcome may underperform in certain situations. Through extensive simulations, the authors demonstrated that such methods can generate more biased estimates than those based on predicting exposure. The theoretical basis for this behavior has already been discussed in detail by authors such as Dominici et al.[26].

The results of this study reinforce the importance of carefully choosing statistical methods when analyzing the association between air pollutants and health outcomes, especially in urban contexts with a high burden of CVD. By comparing approaches based on outcome prediction and exposure prediction, it was shown that the latter strategy offers less biased estimates, strengthening the basis for more robust inferences. These findings contribute to the methodological refinement of environmental epidemiological studies, improving the quality of available scientific evidence.

Besides the technical advancement, this work provides relevant information for the formulation of public policies aimed at mitigating the effects of air pollution on cardiovascular health. By guiding decisions on acceptable air quality standards, environmental monitoring strategies, and the allocation of hospital resources during critical periods, studies like this play a fundamental role in the preventive planning of health systems in large urban centers.

One limitation is the restriction of time-series studies to estimating only acute effects. Chronic effects, on the other hand, are estimated using cohort studies. In this case, this research cannot estimate long-term effects on cardiovascular morbidity. The association measured by time-series data does not necessarily have a causal origin. Cohort studies can infer causality.

Future research should explore approaches that integrate high-resolution spatial and temporal data, analyses stratified by vulnerable groups (such as the elderly or people with comorbidities), and methods that simultaneously consider multiple pollutants and non-linear effects. The incorporation of modern machine learning and causal inference techniques could further expand the predictive and explanatory capacity of models, strengthening the link between science and public policy in environmental health.